\documentclass[reprint,aps,prapplied,superscriptaddress]{revtex4-2}

\usepackage{array}[=2016-10-06]
\usepackage[english]{babel}
\usepackage{amsmath}
\usepackage{graphicx}
\usepackage{float}
\usepackage{physics}
\usepackage{amssymb}
\usepackage[T1]{fontenc}
\usepackage[colorlinks=true, allcolors=blue]{hyperref}
\usepackage{indentfirst}
\usepackage{kotex}
\usepackage{multirow}
\usepackage{xcolor}
\usepackage{upgreek}
\usepackage{textcomp}
\usepackage{array}
\usepackage[utf8]{inputenc}
\usepackage{natbib}
\usepackage{booktabs}

\makeatletter
\def\maketitle{
	\@author@finish
	\title@column\titleblock@produce
	\suppressfloats[t]}
\makeatother 

\begin{document}

\title{
\textbf{Coherent Optical Readout of Neutral-Atom Qubits via Homodyne Detection of Resonant Fluorescence}
}

\author{Minho Jang}
\thanks{These authors contributed equally.}
\affiliation{Department of Physics, Korea Advanced Institute of Science and Technology (KAIST), Daejeon 34141, South Korea}

\author{Jaeyong Hwang}
\thanks{These authors contributed equally}
\altaffiliation{Current address: JILA, NIST and Department of Physics, University of Colorado, Boulder, CO 80309, USA}
\affiliation{Department of Physics, Korea Advanced Institute of Science and Technology (KAIST), Daejeon 34141, South Korea}

\author{Minhyeok Kim}
\affiliation{Department of Physics, Korea Advanced Institute of Science and Technology (KAIST), Daejeon 34141, South Korea}

\author{Sanghyo Park}
\affiliation{Department of Physics, Korea Advanced Institute of Science and Technology (KAIST), Daejeon 34141, South Korea}

\author{Donggyu Kim}
\email{dngkim@kaist.ac.kr}
\affiliation{Department of Physics, Korea Advanced Institute of Science and Technology (KAIST), Daejeon 34141, South Korea}
\affiliation{OQT Inc., Daejeon 34051, South Korea}

\begin{abstract}
Fast, non-destructive qubit readout is essential for repeated sampling, mid-circuit measurement, and fault-tolerant neutral-atom quantum processing, but camera-based fluorescence detection introduces electronic latency beyond fluorescence-collection time. We propose coherent readout of the state-dependent resonant fluorescence using balanced homodyne detection with photodiodes. An analytic model quantifies the effect of atomic motion on the time-integrated quadrature signal and noise, showing that harmonic motion causes phase averaging while photon recoils produce trajectory-dependent signal fluctuations; the competition between signal accumulation and motion-induced decoherence sets an optimal detection time. Three-dimensional Monte Carlo simulations validate the model and predict state-detection fidelities exceeding $99\%$ within $100~\upmu\mathrm{s}$ for experimentally realistic parameters. These results establish the readout conditions and physical constraints for rapid coherent fluorescence readout of neutral-atom qubits.
\end{abstract}

\maketitle

\section{Introduction}

Neutral atoms trapped in optical tweezers have emerged as a leading platform for quantum information processing, combining long-lived internal-state qubits with programmable connectivity~\cite{Saffman2010,Graham2022,Bluvstein2024,Radnaev2024}. Large defect-free arrays assembled through dynamic atom arrangement~\cite{Chiu2025_Nature3000qubit, Bluvstein2026,Bluvstein2024}, together with Rydberg-mediated entangling interactions~\cite{Wilk2010,Jandura2023, Madjarov2020}, provide a route toward scalable quantum processors. Recent experiments have demonstrated fast, high-fidelity optical control~\cite{Evered2023, Evered2026,Bluvstein2024,Radnaev2024,Park2026} and efficient state readout through resonant-fluorescence imaging with single-photon-sensitive cameras~\cite{Graham2022,Bluvstein2024,Yavuz2006,Bochmann2010}.

Practical neutral-atom quantum processors require fast, non-destructive qubit readout. Fault-tolerant operation demands repeated syndrome measurements~\cite{Bluvstein2024,Radnaev2024,Graham2023,Lis2023,Wu2022,Scholl2023,Liu2023,Bluvstein2026}, while quantum simulations and algorithms often rely on repeated sampling of measurement outcomes~\cite{Ebadi2021, Semeghini2021, Ebadi2022, Bluvstein2024}. With recent advances in continuous atom loading~\cite{Chiu2025_Nature3000qubit,Gyger2024,Li2025}, readout latency can become an important constraint on experimental throughput. Although sufficient resonant fluorescence for state discrimination can be collected on the microsecond timescale~\cite{Falconi2025,Su2025,Wang2025,Deist2022}, camera-based detection introduces additional latency through serial pixel readout, electronic amplification, and data transfer~\cite{Bluvstein2024, Norcia2023}. Single-photon-counting detector arrays can reduce this latency~\cite{Fuhrmanek2011,Deist2022,Gibbons2011,Shea2020}, but scaling such detectors to large atom arrays remains technically challenging~\cite{Hao2024, Morimoto2021}.

\begin{figure}[b]
    \centering
    \includegraphics[width=\linewidth]{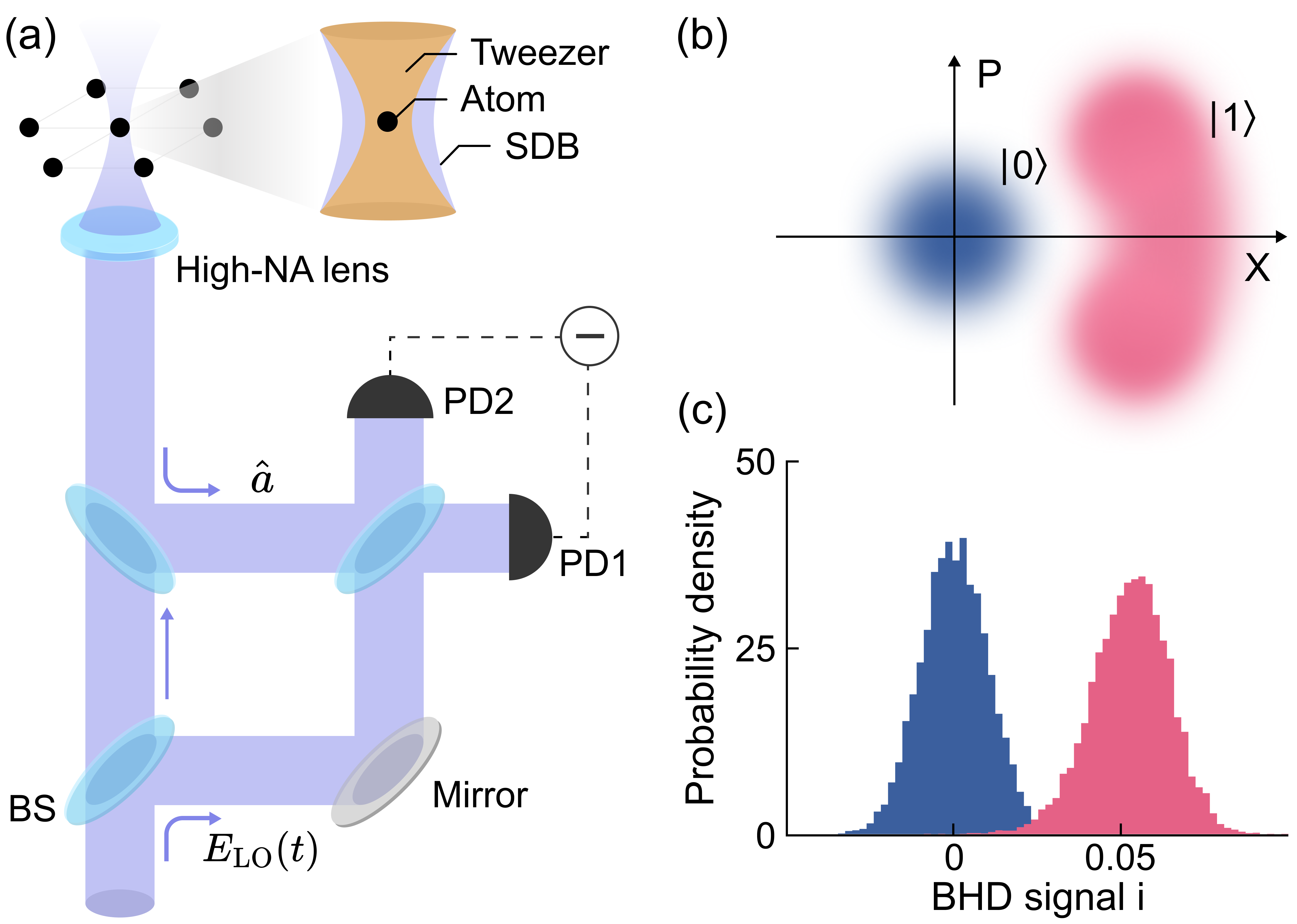}
    \caption{
        \textbf{Balanced homodyne readout of a neutral-atom qubit.}
        \textbf{(a)} Fluorescence from an atom driven by the state-detection beam (SDB) interferes with the local oscillator (LO) $E_\text{LO}(t)$ at a 50:50 beam splitter and is detected by photodiodes PD1 and PD2. $\hat{a}$ denotes the field operator of collected fluorescence mode.
        \textbf{(b)} Phase-space distributions for the optically dark $\ket{0}$ (blue) and bright $\ket{1}$ (red) states, with the bright-state distribution broadened by atomic motion.
        \textbf{(c)} Corresponding $X$-quadrature distributions; the LO phase selects the measured quadrature. 
    }
    \label{fig:1}
\end{figure}

Here, we propose coherent optical readout of neutral-atom qubits using balanced homodyne detection (BHD)~\cite{Gardiner2004,Schumaker1984,Ip2008,Kikuchi2016} of state-dependent resonant fluorescence (Fig.~\ref{fig:1}). Interference with a strong local oscillator converts the weak fluorescence field into a differential quadrature signal measurable with conventional photodiodes, avoiding camera readout latency. Atomic motion, however, modulates the fluorescence phase relative to the local oscillator, while photon scattering heats the atom and produces trajectory-dependent signal fluctuations. We develop an analytic model that relates these motional dynamics to the BHD signal and signal-to-noise ratio. Three-dimensional Monte Carlo simulations validate the model and quantify the resulting state-detection fidelity, revealing an optimal detection time set by the competition between signal accumulation and motion-induced decoherence. For a 7 mK-deep trap and an overall detection efficiency of 7.5\%, the simulations predict state-detection fidelities exceeding 99\% within 100~$\upmu\mathrm{s}$. 

\section{Resonant Fluorescence}
We consider a long-lived ground-state qubit encoded in hyperfine Zeeman sublevels~\cite{Rosenfeld2011} or nuclear-spin states~\cite{Ma2022,Jenkins2022} of an atom confined in an optical tweezer. A state-detection beam (SDB) at angular frequency $\omega_L$ drives a cycling transition between the bright state $\ket{1}$ and an excited state $\ket{e}$ with detuning $\Delta$. The atom--light interaction is characterized by the saturation parameter $s=2\Omega^2/(4\Delta^2+\gamma^2)$, where $\Omega$ is the Rabi frequency and $\gamma$ is the radiative decay rate. Because spatial variations in $\Omega$ and $\Delta$ across the atomic trajectory are negligible under the conditions considered, we treat $s$ as constant. To suppress motional heating from dipole-force fluctuations (DFF), we operate in the far-detuned, weak-saturation regime, $|\Delta|\gg\gamma$ and $s\simeq\Omega^2/2\Delta^2\ll1 $~\cite{Martinez2018,Fuhrmanek2011}.

Under SDB illumination, an atom in the bright state $\ket{1}$ scatters photons on the cycling transition, whereas an atom in $\ket{0}$ remains optically dark. The steady-state collection rate is 
\begin{equation}
    \label{eq:mean_pho}
    R(s)=\eta_\text{c}\frac{\gamma}{2}\frac{s}{1+s}\doteq \eta_\text{c}\Gamma(s),
\end{equation}
where $\Gamma(s)$ is the total photon-scattering rate and $\eta_\text{c}$ is the collection efficiency, including the finite numerical aperture and transmission of the collection optics. The resonant coherent component is collected at the rate 
\begin{equation}
    \label{eq:coh_scat}
    R_\text{rc}(s)=\frac{R(s)}{1+2s},
\end{equation}
which approaches $R(s)$ in the weak-saturation limit~\cite{Tannoudji1998}. The remaining fluorescence consists of an incoherent component,
$R_\text{ic}(s)=R(s)s/(1+s)$, and a nonresonant coherent component associated with the Mollow sidebands,
$R_\text{nrc}(s)=R_\text{rc}(s)s/(1+s)$, such that
$R(s)=R_\text{rc}(s)+R_\text{nrc}(s)+R_\text{ic}(s)$~\cite{Supp}.

\begin{figure}[b]
\centering
\includegraphics[width=\linewidth]{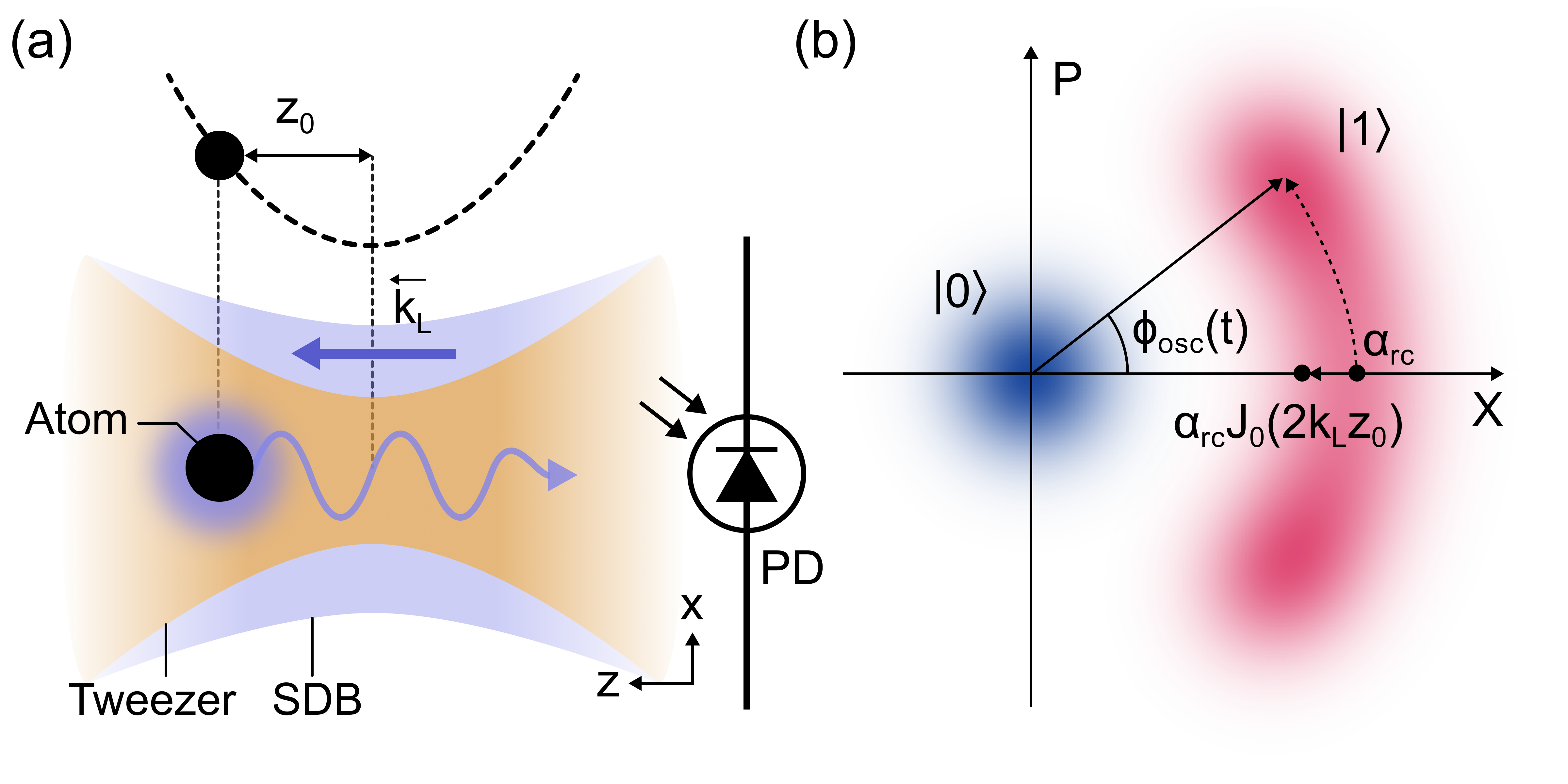}
\caption{
    \textbf{Motion-induced phase averaging.}
    \textbf{(a)} Axial motion produces the fluorescence phase modulation $\phi_{\rm osc}(t)=2k_Lz(t)$. $\vec{k}_L$ denotes the SDB wavevector, and the curved arrow indicates fluorescence collected by the photodiode (PD) in the backward direction, opposite to $\vec{k}_L$. $z_0$ denotes the axial oscillation amplitude in the harmonic approximation.
    \textbf{(b)} Corresponding phase-space trajectory and time-averaged $X$-quadrature signal for fixed-amplitude harmonic motion, reduced from $\alpha_\text{rc}$ to $\alpha_\text{rc} J_0(2k_Lz_0)$.
    }\label{fig:2}
\end{figure}

For each stochastic atomic trajectory, we represent the collected field by an effective complex amplitude $\alpha(s,t)$ in optical phase space,
\begin{align}\label{eq:fluorescence_field}
\begin{split}
    \alpha(s,t)=&\left[\alpha_\text{rc}(s)  +\alpha_\text{nrc}(s,t)+\alpha_\text{ic}(s,t)\right] \\
      &\quad\quad\quad\times \exp[-i(\omega_Lt+\phi_\text{osc}(t))],
\end{split}
\end{align}
with 
\begin{align}\label{eq:field_components}
    \begin{split}
        \alpha_\text{rc}(s) &= \sqrt{R_\text{rc}(s)}, \\
        \alpha_\text{nrc}(s,t) &= \sqrt{2R_\text{nrc}(s)}\cos(\Omega' t), \\
        \alpha_\text{ic}(s,t) &= \sqrt{R_\text{ic}(s)}e^{-i\phi_{\text{rnd}}(t)}.
    \end{split}
\end{align}
The three amplitudes describe the resonant coherent, nonresonant coherent, and incoherent components of the collected fluorescence, respectively. Here, $\Omega'\equiv\sqrt{\Delta^2+\Omega^2}$, and $\phi_\text{osc}(t)\simeq2k_Lz(t)$ is the fluorescence phase shift induced by atomic motion, as illustrated in Fig.~\ref{fig:2}(a). We model the incoherent component with a stochastic phase $\phi_\text{rnd}$ uniformly distributed over $[-\pi,\pi)$. In the far-detuned regime, we neglect the small dispersive phase shift $\delta\phi\sim\arctan[-\gamma/2\Delta]$~\cite{Aljunid2009}. We also neglect motional corrections to the scattering rates since the optical-tweezer trap frequencies are much smaller than the radiative decay rate $\gamma$~\cite{Karan2024}.

\section{Balanced Homodyne Detection of \\ Resonant Fluorescence} \label{sec:BHD}

Balanced homodyne detection (BHD) measures a selected quadrature of the collected fluorescence by interfering it with a strong coherent local oscillator (LO) at a 50:50 beam splitter, as illustrated in Fig.~\ref{fig:1}(a). We denote the fluorescence mode incident on the beam splitter by the field operator $\hat{a}$ and write the LO field as $E_\text{LO}(t)=\text{Re}[\mathcal{E}_0e^{-i\omega_Lt}]$, where $\mathcal{E}_0=E_0e^{i\phi_0}$. The LO phase $\phi_0$ selects the measured fluorescence quadrature. The BHD signal is the differential photocurrent integrated over the detection time $t_\text{d}$. In the strong-LO limit, the corresponding operator is
\begin{align}\label{eq:BHD_operator}
\begin{split}
    \hat{I}_{-} = \int_0^{t_\text{d}} \mathrm{Re}\Big[\sqrt{\eta_\text{d}}\mathcal{E}_0^* e^{i\omega_Lt}\Big(\sqrt{\eta_\text{d}\eta_\text{m}}\hat{a}+\sqrt{1-\eta_\text{d}\eta_\text{m}}\hat{f}\Big)\Big]dt,
\end{split}
\end{align}
where $\text{Re}\big[\hat{\zeta}\big]\equiv (\hat{\zeta}+\hat{\zeta}^\dagger)/2$; $\eta_\text{d}$ is the photodiode quantum efficiency; $\eta_\text{m}$ is the mode-matching efficiency between the resonant fluorescence field and the local oscillator; $\hat{f}$ is a fictitious vacuum mode accounting for optical loss and mode mismatch~\cite{Gardiner2004}. The SDB and LO share a common laser source, making their relative phase largely insensitive to laser-frequency fluctuations and enabling stable phase-sensitive detection. A detailed derivation is provided in Sec.~A of the Supplemental Material~\cite{Supp}. 

Choosing the LO phase to measure the resonant coherent quadrature and evaluating the BHD operator (Eq.~\eqref{eq:BHD_operator}) for the fluorescence field in Eq.~\eqref{eq:fluorescence_field}, the mean BHD signal is
\begin{align}
    \begin{split}
        \langle\hat{I}_{-}\rangle = \eta_\text{d} E_0\sqrt{\eta_\text{m}} \int_0^{t_\text{d}} \alpha_\text{rc}(s)\cos(\phi_\text{osc}(t))dt.
    \end{split}\label{eq:BHD_signal}
\end{align}
The nonresonant coherent component $\alpha_\text{nrc}(s,t)$ averages to zero for detection times long compared with $1/\Omega'$, whereas the incoherent component $\alpha_\text{ic}(s,t)$ vanishes upon averaging over the stochastic phase $\phi_\text{rnd}(t)$. For axial harmonic motion $z(t)=z_0\sin\omega_0 t$, the fluorescence phase relative to the LO is $\phi_\text{osc}(t)=2k_Lz_0\sin\omega_0 t$, where $z_0$ is the oscillation amplitude, $k_L$ is the SDB wavenumber, and $\omega_0$ is the axial trap frequency (Fig.~\ref{fig:2}(a)). For $t_\text{d} \gg 2\pi/\omega_0$, the time-integrated signal and its variance are 
\begin{align}\label{eq10}
\begin{split}
     \langle\hat{I}_{-}\rangle &= \eta_\text{d} E_0t_\text{d}\sqrt{\eta_\text{m}R_\text{rc}(s)}J_0(2k_Lz_0), \\
     \Delta^2\hat{I}_{-} & =  \eta_\text{d} E_0^2t_\text{d}/4,
\end{split} 
\end{align}
where $J_0$ is the zeroth-order Bessel function of the first kind. For harmonic motion with a fixed oscillation amplitude $z_0$, the variance represents the homodyne shot noise. We neglect the additional phase modulation from the Doppler shift because it is much smaller than the position-induced phase modulation $\phi_\text{osc}(t)$ at atomic temperatures of a few hundred microkelvin~\cite{Supp}.

In conventional homodyne detection of amplitude-shift keying (ASK), the bit-error rate is set by the quadrature signal-to-noise ratio (SNR) for a fixed signal amplitude~\cite{Supp}. Here, atomic motion additionally suppresses the time-integrated quadrature signal through the phase-averaging factor $J_0(2k_Lz_0)$, as illustrated in Fig.~\ref{fig:2}(b). The resulting SNR depends on both the detected coherent-photon number and scattering-induced changes in the atomic trajectory during detection. 

\subsection{Heating and Signal-to-Noise Ratio (SNR)}  \label{subsec:SNRinBHD}
\begin{figure}
\centering
\includegraphics[width=\linewidth]{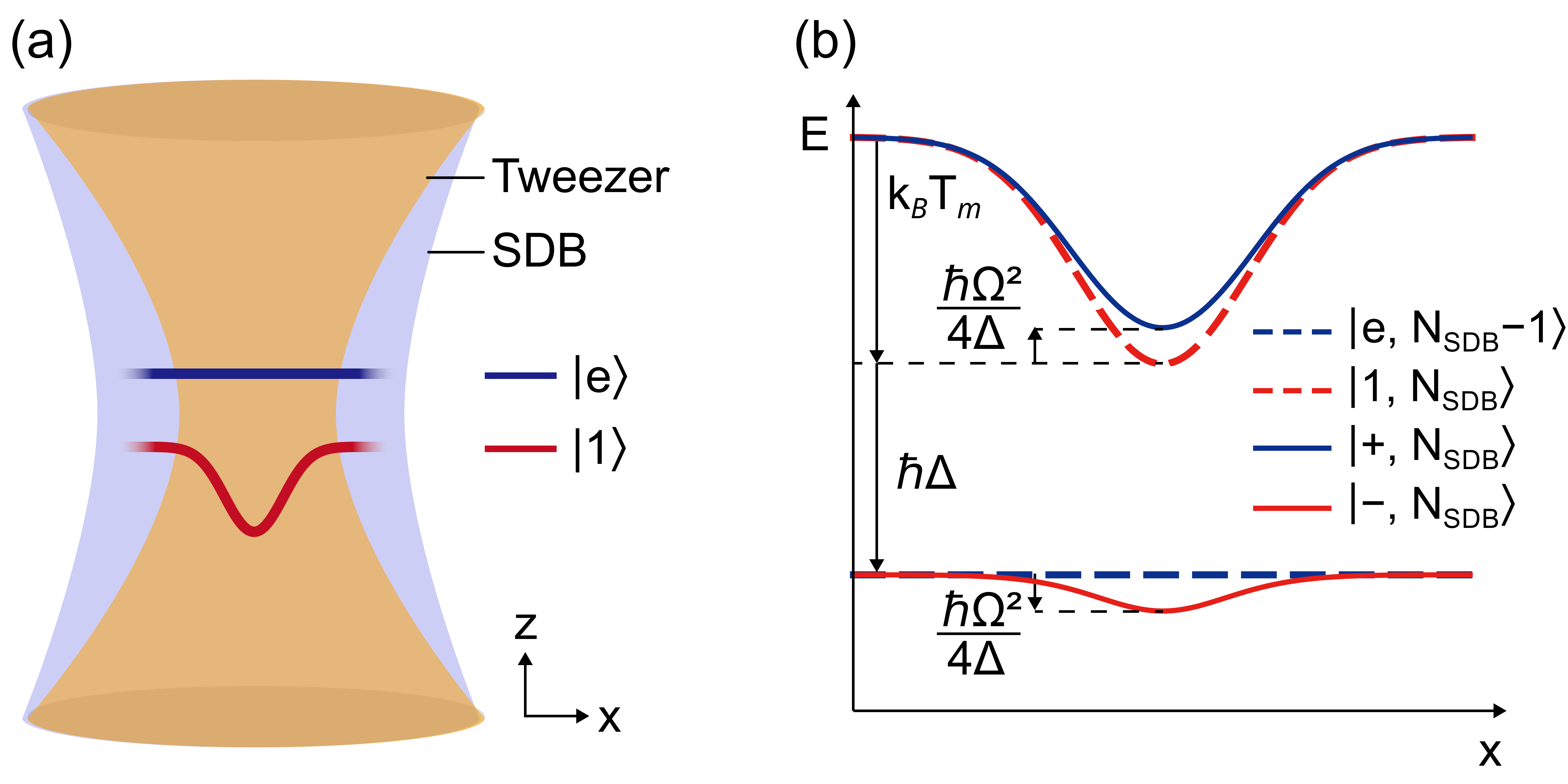}
\caption{
\textbf{Dressed-state energy level diagram.}
    \textbf{(a)} Optical tweezer potentials for $\ket{1}$ and $\ket{e}$ under SDB illumination, with $\ket{e}$ taken to be untrapped. 
    \textbf{(b)} Bare and dressed-state energies in the interaction picture for $\Delta>0$, where $N_\text{SDB}$ is the number of SDB photons.
}
\label{fig:3}
\end{figure}

The phase-averaging factor $J_0(2k_Lz_0)$ assumes harmonic motion with a fixed oscillation amplitude $z_0$. During readout, photon scattering modifies the atomic trajectory through photon recoil and dipole-force fluctuations (DFF), making the oscillation amplitude time dependent. Photon recoil results from the momentum transferred in each scattering event, whereas DFF arises from the different shape of the optical potentials associated with the dressed states
\begin{align}\label{eq:dressed_state}
\begin{split}
   \ket{+,N_\text{SDB}} &= \sin\theta|1,N_\text{SDB}\rangle + \cos\theta|e,N_\text{SDB}-1\rangle, \\
   \ket{-,N_\text{SDB}} &= \cos\theta|1,N_\text{SDB}\rangle - \sin\theta|e,N_\text{SDB}-1\rangle,
\end{split}
\end{align}
where $\theta\equiv\arctan(-\Omega/\Delta)/2 + \pi H(\Delta)/2$, $H$ is the Heaviside step function, and $N_\text{SDB}$ is the number of SDB photons. Figure~\ref{fig:3} illustrates the bare- and dressed-state potentials for blue detuning, $\Delta>0$. Following Ref.~\cite{Martinez2018}, we assume that the excited state $\ket{e}$ experiences no optical-tweezer potential, an approximation shown to reproduce the observed heating dynamics. Photon recoil and the DFF heat the atom, producing trajectory-dependent fluctuations in the oscillation amplitude and fluorescence phase. 

We model the stochastic axial motion as $z(t)=z_0(t)\sin(\omega_0 t - \phi_z(t))$, where both $z_0(t)$ and $\phi_z(t)$ evolve under photon recoil and DFF. Each photon scattering event changes the atomic velocity while the position remains continuous, so the post-event amplitude and phase are jointly determined by the atomic phase-space trajectory. Successive events generate harmonic-motion segments with different oscillation amplitudes. For a given stochastic trajectory, the fixed-amplitude phase-averaging factor $J_0(2k_Lz_0)$ in Eq.~\eqref{eq10} is generalized to the trajectory-averaged factor
\begin{equation}
    \bar{J} = \frac{1}{t_\text{d}} \int_0^{t_\text{d}} J_0\left[2k_L z_0(t)\right]dt.
\label{eq:phaseave}
\end{equation}
Here, $J_0$ accounts for phase averaging of the harmonic motion at a fixed amplitude, whereas the time integral averages this factor over the amplitudes sampled during the stochastic trajectory. 

Repeated measurements sample an ensemble of such stochastic trajectories, giving the mean $\lambda$ and total variance $\sigma^2$ of the resulting BHD outcome distribution as
\begin{align}
\begin{split} 
    \label{eq:BHD_signal_ensemble}
    \lambda / E_0 &\simeq \eta_\text{d}t_\text{d} \sqrt{\eta_\text{m}R_\text{rc}(s)} \mathbb{E}[\bar{J}], \\
    \sigma^2/E_0^2  &\simeq \eta_\text{d} t_\text{d}/4 + \eta_\text{d}^2 t_\text{d}^2\eta_\text{m}R_\text{rc}(s)\mathbb{E}[(\delta\bar{J})^2],
\end{split}
\end{align}
where $\mathbb{E}[\cdot]$ denotes the ensemble average over stochastic trajectories, and $\delta \bar{J}\equiv\bar{J}-\mathbb{E}[\bar{J}]$. The first term in the variance is the homodyne shot noise in Eq.~\eqref{eq10}, while the second is the excess noise from trajectory-to-trajectory fluctuations in $\bar{J}$. Defining the detected coherent-photon number as $N_\text{rc}=\eta_\text{d}R_\text{rc}(s)t_\text{d}$, the bright-state signal-to-noise ratio becomes 
\begin{align}
    \begin{split}
        \lambda / \sigma & \simeq 2\sqrt{\eta_\text{m}N_\text{rc}}\times\frac{ \mathbb{E}[\bar{J}]}{\sqrt{1 + 4\eta_\text{m}N_\text{rc}\mathbb{E}[(\delta\bar{J})^2]}}. \\
    \end{split} \label{eq:SNR}
\end{align}
In the absence of atomic motion, $\mathbb{E}[\bar{J}] =1$ and $\mathbb{E}[(\delta\bar{J})^2]=0$, recovering the standard homodyne scaling $\lambda/\sigma=2\sqrt{\eta_\text{m}N_\text{rc}}$. With atomic motion, extending the detection time increases $N_\text{rc}$, while accumulating scattering-induced heating, which reduces $\mathbb{E}[\bar{J}]$ and increases $\mathbb{E}[(\delta\bar{J})^2]$. The SNR initially increases as coherent signal accumulates and eventually decreases as motion-induced phase averaging and trajectory fluctuations become dominant, producing an optimal detection time.

\section{Monte Carlo Simulations and State-Detection Error}\label{sec:simulation}

We evaluate the BHD outcome statistics using three-dimensional Monte Carlo (MC) simulations sampling thermal initial conditions and stochastic atomic trajectories generated by photon scattering. The simulations follow the semiclassical dressed-state model of Ref.~\cite{Martinez2018}. We consider a single $^{87}$Rb atom trapped in an Gaussian optical tweezer, with the qubit encoded in the $5S_{1/2}$ ground-state manifold as $\ket{0}=\ket{F=1,m_F=1}$ and $\ket{1}=\ket{F=2,m_F=2}$. The tweezer wavelength and magnetic-field configuration follow Ref.~\cite{Martinez2017}, and the initial atomic temperature is $T=30~\upmu\mathrm{K}$.

The SDB is far detuned from the cycling transition between $\ket{1}$ and $\ket{e}=\ket{5P_{3/2},F'=3,m_F'=3}$. Unless otherwise stated, the simulations use $s=0.08$ and $\Delta=2\pi\times123$ MHz at the SDB center. These parameters produce an average of approximately $270$ scattering events per trajectory during $t_\text{d}=200~\upmu\mathrm{s}$, corresponding to a mean motional energy increase of approximately $k_\mathrm{B}\times0.1~\mathrm{mK}$ across the trajectories. The atom-loss probability is $0.034\%$ for 1 mK-deep trap, whereas no loss events are observed among 50,000 trajectories for a trap depth of 3 mK or greater. The MC simulations use an overall detection efficiency of $\eta\equiv\eta_\text{c}\eta_\text{m}\eta_\text{d}=7.5\%$, comprising a photodiode quantum efficiency $\eta_\text{d}=90\%$, a mode overlap efficiency $\eta_\text{m}=80\%$ and a collection efficiency $\eta_\text{c}=10.3\%$. The collection efficiency corresponds to $18.7\%$ geometric collection by a 0.7-NA objective and approximately $55\%$ optical transmission. The simulation parameters are summarized in Table~\ref{table:simulation_parameters}, and the MC procedure is described in Sec.~B of the Supplemental Material~\cite{Supp}.

\subsection{BHD Outcome Statistics}\label{subsec:BHD_outcome}

\begin{figure}
    \centering
    \includegraphics[width=\linewidth]{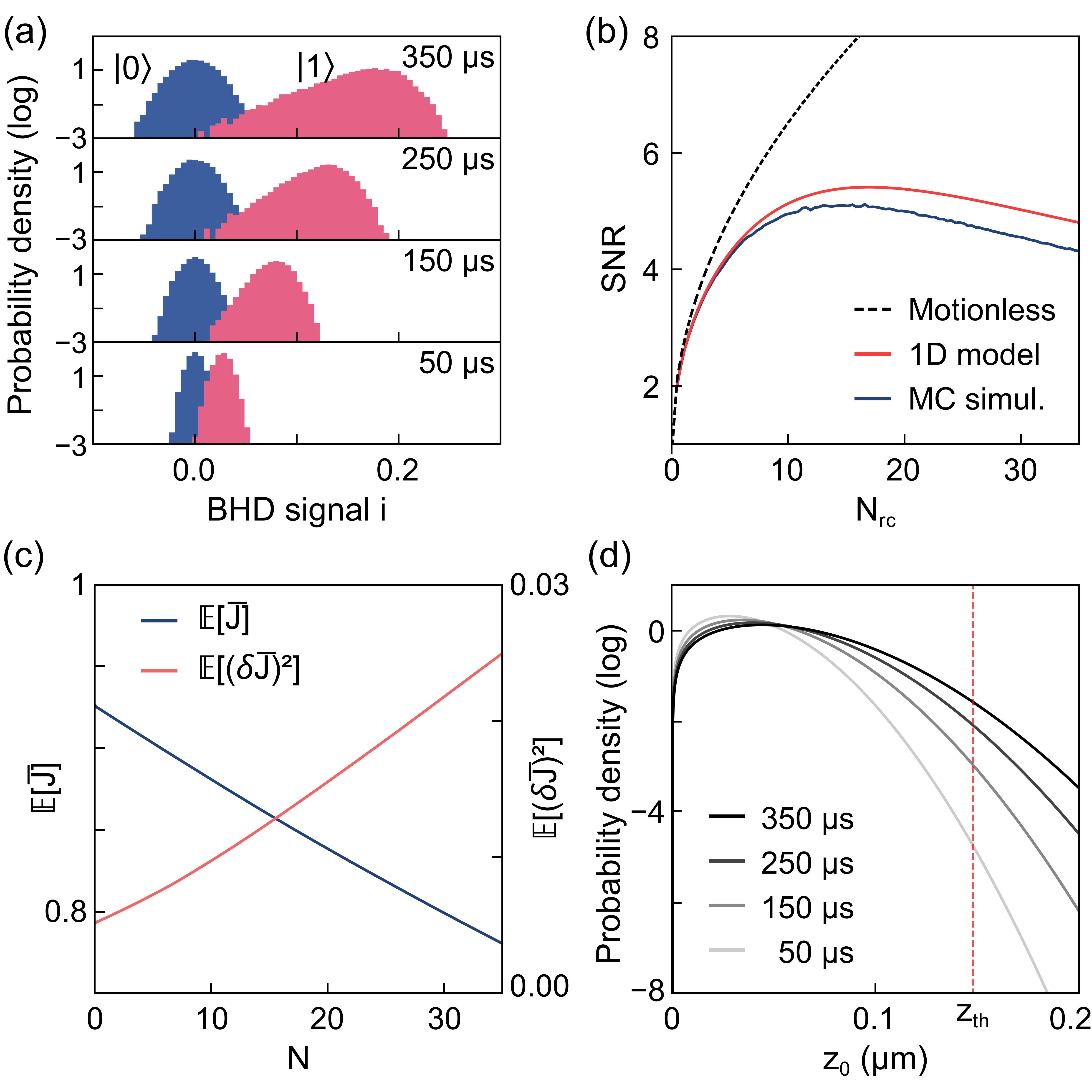}
    \caption{
        \textbf{BHD outcome statistics from Monte Carlo simulations.}
        \textbf{(a)} Conditional BHD outcome distributions $p(i|0)$ (blue) and $p(i|1)$ (red) at selected detection times. 
        \textbf{(b)} Bright-state SNR obtained from the Monte Carlo (MC) simulations and the analytic model (red) in Eq.~\eqref{eq:SNR}. The black dotted line shows the motionless-atom limit, $\mathrm{SNR}=2\sqrt{\eta_\text{m}N_\text{rc}}$, where $N_\text{rc}$ is the detected resonant coherent-photon number.
        \textbf{(c)} Mean phase-averaging factor $\mathbb{E}[\bar{J}]$ and trajectory-induced variance $\mathbb{E}[(\delta\bar{J})^2]$ as functions of the detected photon number $N$. Their offsets at $N=0$ reflect initial thermal motion and variations across the thermal ensemble.
        \textbf{(d)} Distribution of the oscillation amplitude $z_0$ at selected detection times. The dashed line marks $z_\text{th}$ defined by $J_0(2k_Lz_\text{th})=0$. In \textbf{(b)} and \textbf{(c)}, statistical uncertainties, evaluated as one standard deviation across 10 independent ensembles of 5,000 trajectories each, are comparable to the line widths.
        }
    \label{fig:4}
\end{figure}

We first consider the BHD outcome statistics for a plane-wave SDB with uniform intensity and polarization across the atomic trajectory, excluding focused-beam effects. For atoms prepared in $\ket{0}$ and $\ket{1}$, the MC simulations compute the conditional distributions $p(i|0)$ and $p(i|1)$, respectively, as functions of detection time $t_\text{d}$, where $i$ denotes the measurement outcome of the time-integrated BHD operator $\hat{I}_-$ defined in Eq.~\eqref{eq:BHD_operator}. Figure~\ref{fig:4}(a) shows the two distributions $p(i|0)$ and $p(i|1)$ at selected detection times. For the dark state, $p(i|0)$ remains centered at zero, while its width is set by homodyne shot noise, consistent with Eq.~\eqref{eq10}. 

By contrast, $p(i|1)$ reflects the competition between coherent-signal accumulation and motion-induced decoherence as the detection time increases. Figure~\ref{fig:4}(b) plots the SNR extracted from the mean and variance of $p(i|1)$. At short times, $p(i|1)$ shifts toward larger BHD outcomes as the detected resonant coherent-photon number $N_\text{rc}$ increases, approaching the weak-heating scaling SNR $\sim 2\sqrt{\eta_\text{m}N_\text{rc}}$ in Eq.~\eqref{eq:SNR}. At longer times, the peak continues to shift toward larger outcomes, while accumulated scattering-induced heating produces an increasingly long low-signal tail and reduces the SNR. 

The time-dependent BHD outcome statistics are largely accounted for by the recoil-induced growth of the axial oscillation amplitude. We model recoil- and DFF-induced changes in the axial oscillation amplitude $z_0(t)$ and phase $\phi_z(t)$ as a one-dimensional random walk~\cite{Supp}, using the harmonic-motion model between scattering events (Sec.~\ref{subsec:SNRinBHD}). From the sampled trajectories, we evaluate $\mathbb{E}[\bar{J}]$ and $\mathbb{E}[(\delta\bar{J})^2]$, shown in Fig.~\ref{fig:4}(c), and use them in Eq.~\eqref{eq:SNR}. The resulting SNR closely reproduces the three-dimensional MC results (Fig.~\ref{fig:4}(b)). The remaining difference is due to the motional dynamics not captured by the harmonic random-walk model. In the full MC simulation, both axial and radial displacements reduce the effective axial confinement relative to the harmonic approximation, allowing recoil heating to drive larger oscillation amplitudes. The SDB-induced dressed-state potentials further reduce the restoring force through admixture with the untrapped excited state. The resulting motion decreases $\mathbb{E}[\bar{J}]$ and increases $\mathbb{E}[(\delta\bar{J})^2]$, yielding a lower SNR than the one-dimensional harmonic model.

The one-dimensional model also provides a direct interpretation of the low-signal tail in $p(i|1)$. Figure~\ref{fig:4}(d) shows that scattering-induced heating broadens the distribution of $z_0$, increasing the fraction of trajectories approaching or exceeding the characteristic amplitude $z_{\text{th}}$, defined by the first positive root of $J_0(2k_L z_{\text{th}})=0$. For a 780 nm SDB, $z_\text{th}\simeq149~\mathrm{nm}$. Trajectories near $z_{\text{th}}$ produce strongly suppressed time-integrated BHD signals, while those beyond $z_{\text{th}}$ can contribute with the opposite quadrature sign. The resulting cancellation of opposite-sign quadrature contributions produces the low-signal tail in $p(i|1)$. 

\subsection{Detection Error and Optimal Conditions}\label{subsec:motion_decoherence}

\begin{figure}
    \centering
    \includegraphics[width=\linewidth]{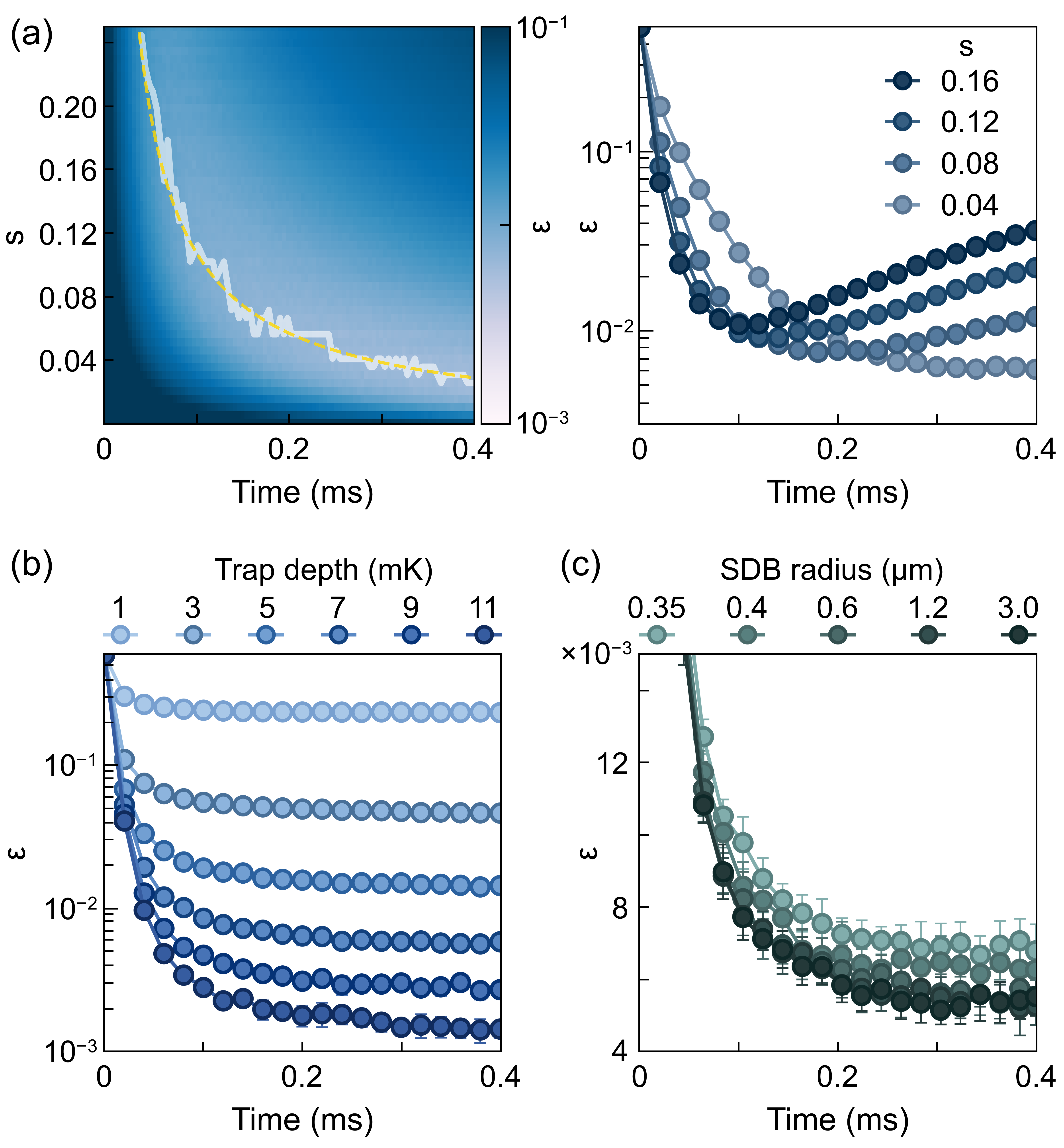}
    \caption{
        \textbf{Dependence of BHD detection error $\epsilon$ on readout conditions.}
        \textbf{(a)} Detection error as a function of the saturation parameter $s$ and detection time $t_\text{d}$ (left), with cross sections for selected $s$ (right). The white contour traces the optimal detection condition, with its width indicating the statistical uncertainty. The yellow dashed line indicates $t_\text{d}\propto1/s$.
        \textbf{(b)} Detection error as a function of detection time $t_\text{d}$ for several optical-tweezer trap depths.
        \textbf{(c)} Detection error as a function of detection time for several SDB $1/e^2$ intensity radii in a 7 mK-deep trap.
        Error bars indicate one standard deviation across 10 independent ensembles, each containing 5,000 trajectories.
        For \textbf{(b)} and \textbf{(c)}, $s$ is chosen to minimize the detection error at each detection time, trap depth and SDB radius. 
    }
    \label{fig:5}
\end{figure}

We discriminate the qubit state by assigning BHD outcomes below a threshold $i_\text{th}$ to the dark state and those at or above the threshold to the bright state. The corresponding misidentification probabilities are $\epsilon_\mathrm{0}=\sum_{i\geq i_\text{th}}p(i|0)$ and $\epsilon_\mathrm{1}=\sum_{i < i_\text{th}}p(i|1)$~\cite{Dassonneville2020}. The threshold $i_\mathrm{th}$ is chosen to minimize the average detection error $\epsilon=\frac{1}{2}\left(\epsilon_\mathrm{0}+\epsilon_\mathrm{1}\right)$ with $p(i_\text{th}|0)=p(i_\text{th}|1)$. The detection fidelity is defined as $F\equiv 1-\epsilon$, and is determined by the overlap between $p(i|0)$ and $p(i|1)$. For a fixed saturation parameter $s$,  increasing the detection time initially reduces $\epsilon$ as the coherent signal accumulates (Fig.~\ref{fig:5}(a)). At longer times, the low-signal tail of $p(i|1)$ increases its overlap with $p(i|0)$, causing the error to increase and producing an optimal detection time. 

The optimal detection condition balances the detected coherent-photon number required for state discrimination against motion-induced decoherence accumulated through photon scattering. In the weak-saturation regime, the optimal detection time scales approximately as $1/s$, while the detected photon number at the optimum  remains nearly constant at $\sim21$ over the range of $s$ considered in Fig.~\ref{fig:5}(a). Increasing $s$ reduces the optimal detection time but increases the minimum error because the resonant coherent fraction of the fluorescence decreases as $R_\text{rc}(s) / R(s) = 1/(1+2s)$. Lower saturation reduces the minimum error at the cost of a longer detection time. Deeper traps provide stronger confinement, further lowering the minimum error, as shown in Fig.~\ref{fig:5}(b).

We next consider site-selective readout with a tightly focused SDB, which can reduce the need to transport atoms to a separate readout region in large arrays~\cite{Martinez2017, Radnaev2024}. Tight focusing introduces longitudinal polarization components that can drive unwanted transitions out of the cycling manifold (see Fig.~\ref{fig:S2}(b) in Supplemental Material~\cite{Supp}). We incorporate the spatial polarization profile into the Monte Carlo simulations and calculate the detection error for a 7 mK-deep trap as a function of the SDB $1/e^2$ intensity radius. As shown in Fig.~\ref{fig:5}(c), smaller beam radii increase the detection error because stronger polarization mixing enhances population leakage out of the cycling manifold and reduces the coherent fluorescence signal. With $s$ optimized to minimize the detection error at each detection time and SDB radius, the simulations predict detection fidelities exceeding $99\%$ within $100~\upmu\mathrm{s}$ for SDB radii from $0.6$ to $3.0~\upmu\mathrm{m}$.

\section{Discussion}

\begin{figure}
\includegraphics[width=\linewidth]{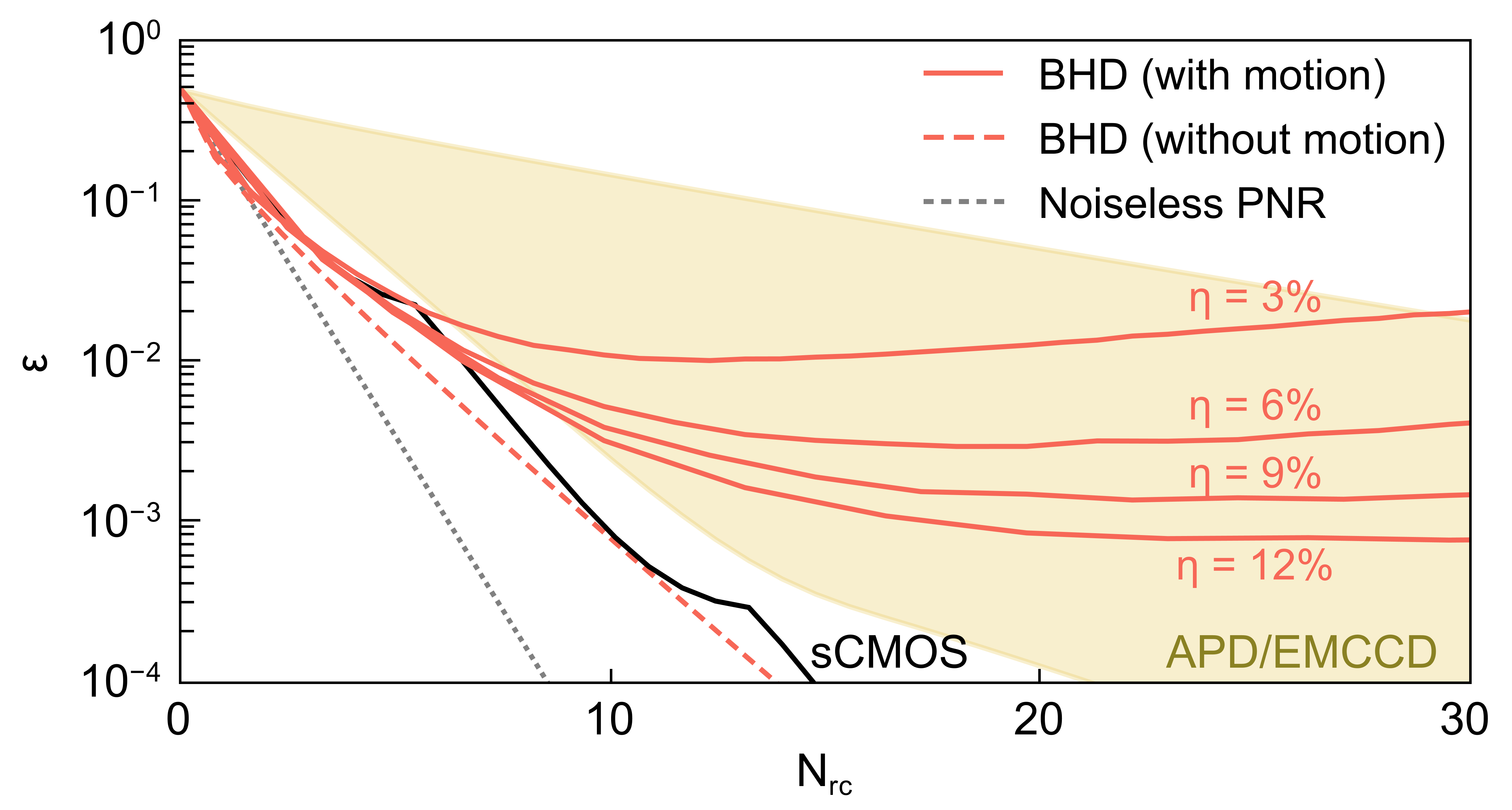}
\caption{
    \textbf{Comparison of neutral-atom qubit readout errors.}
     BHD with and without atomic motion (solid and dashed red lines, respectively) is compared with noiseless photon-number-resolving (PNR) detection (black dashed line) for different detection efficiencies $\eta$ as a function of detected coherent-photon number $N_\text{rc}$. The gray curve and yellow shaded region show  detection errors for modeled sCMOS and APD/EMCCD. Detector-noise models and parameters are provided in Table~\ref{table2} in Supplemental Material~\cite{Supp}. 
    }
\label{fig:6}
\end{figure}

We benchmark coherent fluorescence readout against noiseless photon-number-resolving (PNR) detection and camera-based fluorescence detection under the same photon-collection conditions (Fig.~\ref{fig:6}). BHD achieves comparable detection errors while replacing serial camera readout with fast differential quadrature measurement using photodiodes. Because BHD is sensitive to fluorescence phase, scattering-induced motion limits the detection fidelity. Increasing the detection efficiency $\eta$ reduces the detection error by increasing the detected coherent signal without additional photon scattering, and additional axial confinement can further suppress motion-induced phase averaging by reducing the oscillation amplitude.

The detection fidelity may be further improved by reducing the quadrature noise or increasing the separation between the state-dependent BHD signals. Quadrature squeezing by a factor $e^{-r}$ reduces the homodyne noise and, in the weak-heating regime, is equivalent to increasing the detected coherent-photon number from $N_\text{rc}$ to $N_\text{rc}e^{2r}$~\cite{Slusher1990,Wu1986,Grangier1987,Heidmann1987}. Alternatively, a binary phase-shift keying (BPSK) scheme~\cite{Ip2008,Kikuchi2016} can assign opposite fluorescence phases to $\ket{0}$ and $\ket{1}$ using two state-selective detection fields with a relative phase of $\pi$. Accounting for the shot noise from two LOs, the increased quadrature separation is equivalent to increasing the effective photon number from  $N_\text{rc}$ to $2N_\text{rc}$. 

\section{Conclusion}
We have developed a theoretical framework for coherent optical readout of neutral-atom qubits through balanced homodyne detection of resonant fluorescence. The analytic model shows that harmonic motion suppresses the time-integrated quadrature signal, while scattering-induced trajectory fluctuations reduce its mean and introduce excess noise. Three-dimensional Monte Carlo simulations validate these dynamics and quantify the resulting state-detection fidelity. The competition between coherent-signal accumulation and motion-induced decoherence sets an optimal detection time. For a 7 mK-deep trap and an overall detection efficiency of 7.5\%, the simulations predict state-detection fidelities exceeding $99\%$ within $100~\upmu\mathrm{s}$. Stronger confinement suppresses motion-induced decoherence, while higher detection efficiency allows sufficient signal to be collected with fewer scattering events. 

Coherent fluorescence readout provides an alternative to serial camera readout through differential quadrature detection with fast photodiodes while maintaining comparable detection errors. Integration with multichannel balanced receivers could extend this approach to a parallel readout of large atom arrays. Reduced readout latency could benefit repeated syndrome measurements and high-throughput sampling in neutral-atom quantum processors.

\bibliography{references}

\clearpage

\setcounter{equation}{0}
\setcounter{figure}{0}
\setcounter{table}{0}
\setcounter{page}{1}

\renewcommand{\theequation}{S\arabic{equation}}
\renewcommand{\thefigure}{S\arabic{figure}}
\renewcommand{\thetable}{S\arabic{table}}
\renewcommand{\bibnumfmt}[1]{[S#1]}

\title{
\textbf{\large Supplemental Material of ``Coherent Optical Readout of Neutral-Atom Qubits via Homodyne Detection of Resonant Fluorescence''}
}

\maketitle
\onecolumngrid

\subsection{Coherent detection signal derivation}\label{supp:signal_derivation}
To derive the balanced homodyne detection (BHD) signal, we consider an ideal 50:50 beam splitter (BS) that mixes the collected fluorescence field operator $\hat{a}$ with a strong coherent local oscillator (LO)  $E_{\text{LO}}(t)=\mathrm{Re}[\mathcal{E}_0e^{-i\omega_L t}]$, where the complex amplitude is defined as $\mathcal{E}_0=E_0e^{i\phi_0}$, with $E_0$ real. In the strong-LO limit, the LO is treated as a classical coherent field. The two output field operators of the BS, $\hat{a}_1$ and $\hat{a}_2$, are
\begin{equation}\label{seq:a1a2}
    \hat{a}_1 =\frac{1}{\sqrt{2}}\left[\mathcal{E}_0e^{-i\omega_L t}+i\hat{a}\right],\quad
    \hat{a}_2 =\frac{1}{\sqrt{2}}\left[i\mathcal{E}_0e^{-i\omega_L t}+\hat{a}\right]. \\
\end{equation}

The photocurrent measured at each photodetector is proportional to the photon number operator $\hat{n}'_i = \hat{a}{'}_i^\dagger \hat{a}{'}_i$ ($i = 1, 2$). Finite detection efficiency is modeled as
\begin{equation}
    \hat{a}'_i=\sqrt{\eta_\text{d}}\hat{a}_i+\sqrt{1-\eta_\text{d}}\hat{f}_i,
\end{equation}
where $\eta_\text{d}$ is the photodetector quantum efficiency and $\hat{f}_i$ is a fictitious vacuum mode accounting for optical loss~\cite{Militaru2022, Leonhardt1993}. To include imperfect spatial and polarization matching, we define the power mode matching efficiency as
\begin{equation}
    \eta_\text{m}=\frac{|\int\mathbf{u}^*_\text{LO}(\mathbf{r})\cdot \mathbf{u}_\alpha(\mathbf{r})dA|^2}{(\int|\mathbf{u}_\text{LO}(\mathbf{r})|^2dA)(\int|\mathbf{u}_\alpha(\mathbf{r})|^2dA)},
\end{equation}
where $\mathbf{u}_\text{LO}$ and $\mathbf{u}_\alpha$ are the LO and collected-fluorescence mode profiles, respectively. We replace the fluorescence operator in Eq.~\eqref{seq:a1a2} by
\begin{equation}
    \hat{a}\rightarrow \sqrt{\eta_\text{m}}\hat{a}+\sqrt{1-\eta_\text{m}}\hat{f}_\text{m},
\end{equation}
where $\hat{f}_\text{m}$ is a fictitious vacuum mode for mode mismatch. Its vacuum contribution is combined with the detector loss modes into the effective vacuum mode $\hat{f}$ as
\begin{equation}
    \hat{f}=\frac{\sqrt{1-\eta_\text{d}}(\frac{\hat{f}_2-i\hat{f}_1}{\sqrt2})+\sqrt{\eta_\text{d}(1-\eta_\text{m})}\hat{f}_\text{m}}{\sqrt{1-\eta_\text{d}\eta_\text{m}}}.
\end{equation}

Subtracting the two photocurrents and integrating over the detection time $t_\text{d}$ gives 
\begin{align}
\begin{split}
    \hat{I}_{-} &\equiv \int_0^{t_\text{d}}\frac{1}{2}\left(\hat{n}'_1-\hat{n}'_2\right)dt \\
    &= \int_0^{t_\text{d}} \mathrm{Re}\Big[\sqrt{\eta_\text{d}}\mathcal{E}_0^* e^{i\omega_Lt}\Big(\sqrt{\eta_\text{d}\eta_\text{m}}\hat{a}+\sqrt{1-\eta_\text{d}\eta_\text{m}}\hat{f}\Big)\Big]dt, \\
    &= \int_0^{t_\text{d}} \left[\eta_\text{d}\sqrt{\eta_\text{m}}(\mathcal{E}_0^*e^{i\omega_Lt}\hat{a}+\mathcal{E}_0e^{-i\omega_Lt}\hat{a}^\dagger)/2+\sqrt{\eta_\text{d}(1-\eta_\text{d}\eta_\text{m})}(\mathcal{E}_0^*e^{i\omega_Lt}\hat{f}+\mathcal{E}_0e^{-i\omega_Lt}\hat{f}^\dagger)/2\right]dt,\\
\end{split}
\end{align}
where $\pi/2$ phase is absorbed into $\phi_0$ for convenience. In obtaining this expression, we use the strong-LO approximation, $|E_0|^2 \gg \langle\hat{a}^\dagger\hat{a}\rangle$, and neglect terms that do not scale with the LO amplitude. 

Choosing the LO phase $\phi_0$ to measure the desired quadrature and defining the Fourier components $\tilde{a}(\omega)\equiv \frac{1}{t_\text{d}}\int_0^{t_\text{d}}\hat{a}e^{i\omega t}dt$ for  $t_\text{d}\gg2\pi/\omega_L$, and analogously $\tilde{f}(\omega)$, the BHD operator is written as
\begin{equation}
    \hat{I}_{-} = E_0 t_\text{d}\Big[\eta_\text{d}\sqrt{\eta_\text{m}}\left(\tilde{a}(\omega_L) + \tilde{a}^\dagger(\omega_L)\right)/2+\sqrt{\eta_\text{d}(1-\eta_\text{d}\eta_\text{m})}(\tilde{f}(\omega_L)+\tilde{f}^\dagger(\omega_L))/2 \Big].  \\
\end{equation}
Using the commutator relation $[\hat{\zeta}(t),\hat{\zeta}^\dagger(t')]=\delta(t-t')$~\cite{Gardiner2004}, we obtain
\begin{equation}
    [\tilde{\zeta}(\omega_L),\tilde{\zeta}^\dagger(\omega_L)] = \frac{1}{t_\text{d}}, \\
\end{equation}
for $\zeta=a,f$. The mean BHD signal for a given atomic trajectory is
\begin{align}
\begin{split}
    \langle \hat{I}_- \rangle &= \eta_\text{d} E_0t_\text{d}\sqrt{\eta_\text{m}}\langle \text{Re}(\tilde{a}(\omega_L)) \rangle.
\end{split}\label{eq:S6}
\end{align}
Here, $\langle\cdots\rangle$ denotes the expectation value over the
fluorescence field for a given atomic trajectory.

First, we consider axial harmonic motion with a fixed amplitude $z_0$ and trap frequency $\omega_0$. The corresponding coherent fluorescence amplitude is
\begin{equation}
\alpha(t) = \alpha_\text{rc}(s)\exp\left(-i\omega_L t-i 2k_Lz_0\sin(\omega_0t)\right). \\
\end{equation}
For this fixed trajectory, the mean and variance of the BHD signal $\hat{I}_-$ are given by Eq.~\eqref{eq10} of the main text:
\begin{align}
\begin{split}
    \langle \hat{I}_- \rangle &= \eta_\text{d} E_0t_\text{d} \sqrt{\eta_\text{m}R_\text{rc}(s)}J_0(2k_Lz_0),  \\
    \Delta^2\hat{I}_{-} &= \langle\hat{I}^2_{-}\rangle-\langle\hat{I}_{-}\rangle^2= \eta_\text{d} E_0^2t_\text{d}/4,  \\
\end{split}
\end{align}
where $J_0$ is the zeroth-order Bessel function. 

We next include motional heating by considering the coherent fluorescence amplitude along a given stochastic trajectory, 
\begin{equation}
\alpha(t) = \alpha_\text{rc}(s)\exp\big[-i\omega_L t-i 2k_Lz_0(t)\sin(\omega_0t-\phi_z(t))\big].  \\
\end{equation}
Suppose recoil events occur at times $t:[t_1,t_2,\cdots,t_n]$ during detection time $0\leq t\leq t_\text{d}$. These events divide the detection window into $n+1$ intervals, with $t_0=0, t_{n+1}=t_\text{d}$, and $\Delta t^{(i)}\equiv t_{i+1}-t_i$. Between successive events, the atom undergoes harmonic motion with fixed amplitude and phase, 
\begin{align}
\begin{split}
    z_0(t) &= z_0^{(i)}, ~~~~~ \text{for } t_i\leq t < t_{i+1}\\
    \phi_z(t) &= \phi_z^{(i)}, ~~~~~ \text{for } t_i\leq t < t_{i+1}\\
\end{split}
\end{align}
with $i=0,1,\cdots,n$, for $t_0 = 0$ and $t_{n+1}=t_\text{d}$. 
The mean BHD signal $\langle \hat{I}_- \rangle$ accumulated over $t_\text{d}$ is then
\begin{align}
\begin{split}
    \langle \hat{I}_- \rangle &= \eta_\text{d}E_0 \sqrt{\eta_\text{m}R_\text{rc}(s)} \int_0^{t_\text{d}} \cos\left( 2k_Lz_0(t)\sin(\omega_0t-\phi_z(t)) \right) dt \\
    &= \eta_\text{d}E_0 \sqrt{\eta_\text{m}R_\text{rc}(s)} \sum_{i=0}^{n} \int_{t_i}^{t_{i+1}} \cos\left( 2k_Lz_0^{(i)} \sin(\omega_0t-\phi_z^{(i)}) \right) dt \\
    & \simeq\eta_\text{d} E_0\sqrt{\eta_\text{m}R_\text{rc}(s)}\sum_{i=0}^{n}\Delta t^{(i)}J_0\!\left(2k_Lz_0^{(i)}\right)\\
    &=\eta_\text{d} E_0\sqrt{\eta_\text{m}R_\text{rc}(s)}\int_0^{t_\text{d}}dt
    J_0\!\left(2k_Lz_0(t)\right) = \eta_\text{d} E_0\sqrt{\eta_\text{m}R_\text{rc}(s)}t_\text{d}\bar{J}.
\end{split}\label{eq:S11}
\end{align}
The approximation in Eq.~\eqref{eq:S11} can be understood using the Jacobi--Anger expansion, which decomposes the mean BHD signal $\langle \hat{I}_- \rangle$ into
\begin{equation}
    \langle \hat{I}_- \rangle =\langle \hat{I}_- \rangle_{0}+\sum_{m=1}^{\infty}\langle \hat{I}_- \rangle_{2m}, 
\end{equation}
where
\begin{align}
\begin{split}
    \langle \hat{I}_- \rangle_{0} &= \eta_\text{d} E_0\sqrt{\eta_\text{m}R_\text{rc}(s)} \times \int_0^{t_\text{d}}J_0(2k_Lz_0(t))dt,\\
    \langle \hat{I}_- \rangle_{2m} &= \eta_\text{d} E_0\sqrt{\eta_\text{m}R_\text{rc}(s)} \times \int_0^{t_\text{d}} 2J_{2m}(2k_Lz_0(t))\cos[2m(\omega_0t-\phi_z(t))]dt.
\end{split}
\end{align}
When $z_0(t)$ and $\phi_z(t)$ vary slowly compared with $2\pi/\omega_0$ (i.e., the changes in $z_0^{(i)}$ and $\phi_z^{(i)}$ are sufficiently small), the higher-order terms oscillate rapidly and are strongly suppressed upon time integration, leaving the $\langle \hat{I}_- \rangle_0$ contribution as the dominant term. For detection times $t_\text{d}\gg2\pi/\omega_0$, we verify the contribution of the higher-order terms using the 1D random-walk model. At $s=0.08$, the ratio $|\sum_{m=1}^{\infty}\langle\hat{I}_-\rangle_{2m}/\langle\hat{I}_-\rangle_0|$ remains below $8\times10^{-3}$ for all 50,000 trajectories at detection time $t_\text{d}\geq50~\upmu\mathrm{s}$, supporting the neglect of $\langle\hat{I}_{-}\rangle_{2m}$ terms.

Repeated measurements sample an ensemble of stochastic trajectories, for which $\bar{J}$ is a trajectory-dependent random variable. For a given trajectory, the conditional mean BHD signal is 
\begin{equation}
    \langle \hat{I}_- \rangle_{\bar{J}} \simeq \eta_\text{d}E_0 t_\text{d} \sqrt{\eta_\text{m}R_\text{rc}(s)} \bar{J},
\end{equation}
while the homodyne shot-noise variance remains $\eta_\text{d}E_0^2t_\text{d}/4$. Across the trajectory ensemble, fluctuations in $\bar{J}$ introduce an additional ensemble variance. Denoting the mean and total variance of the resulting BHD outcome distribution by $\lambda$ and $\sigma^2$, we obtain 
\begin{align}
\begin{split} 
    \lambda / E_0 &\simeq \eta_\text{d}t_\text{d} \sqrt{\eta_\text{m}R_\text{rc}(s)} \mathbb{E}[\bar{J}], \\
    \sigma^2/E_0^2  &\simeq \eta_\text{d} t_\text{d}/4 + \eta_\text{d}^2 t_\text{d}^2\eta_\text{m}R_\text{rc}(s)\mathbb{E}[(\delta\bar{J})^2],
\end{split}
\end{align}
where $\mathbb{E}[\cdot]$ denotes the ensemble average over stochastic trajectories, and $\delta \bar{J}\equiv\bar{J}-\mathbb{E}[\bar{J}]$. This reproduces Eq.~\eqref{eq:BHD_signal_ensemble}.

\subsection{Monte Carlo simulations of atomic motion} \label{supp:MC_simulation}
We simulate the atomic motion using Monte Carlo trajectories sampled over thermal distributions of initial position and momentum. Specifically, we implement the dressed-state dipole-force-fluctuation (DFF) model of Ref.~\cite{Martinez2018} for a thermal atom in an optical tweezer under a far-detuned detection beam. In this semiclassical approximation, the atomic center-of-mass motion is treated classically, while internal-state transitions are treated quantum mechanically. The semiclassical approximation is applicable when the photon-scattering rate exceeds the recoil frequency, $\Gamma(s)\gg\omega_{\mathrm{rec}}=\frac{\hbar k_L^2}{2m},$ providing a separation between the internal-state and recoil timescales~\cite{Dalibard1985_motion}. For the parameters considered here, $\Gamma(s)/\omega_{\mathrm{rec}}\simeq60$ at $s=0.08$, and the ratio increases with $s$.

Within this framework, atomic motion is represented by a stochastic trajectory driven by discrete photon recoils and transitions between the dressed-state potentials. The corresponding heating rate, including both photon recoil and DFF, can be evaluated analytically and agrees with the Monte Carlo simulations. The resulting growth of the motional amplitude decreases $\mathbb{E}[\bar J]$ and increases $\mathbb{E}[(\delta\bar J)^2]$. Accordingly, the SNR $\lambda/\sigma$ initially increases as $\sqrt{t_\text{d}}$ in the weak-heating regime and eventually decreases as motion-induced phase averaging and trajectory fluctuations become dominant.

The following subsections present (1) the dressed-state formulation and transition rates used in the main text; (2) a one-dimensional random-walk model and analytical estimates of heating from photon recoil and DFF; (3) the spatial and parameter dependence of the SDB; and (4) the full Monte Carlo simulation procedure.

\subsubsection{Dressed-state picture}\label{supp:dressed_state}
The transition rates between the $N$-photon manifold $\{ \ket{+,N},\ket{-,N} \}$ and the $(N-1)$-photon manifold $\{ \ket{+,N-1}, \ket{-,N-1}\}$ are~\cite{Tannoudji1985}
\begin{align}
\begin{split}
    \Gamma_{+\rightarrow+} = \Gamma_{-\rightarrow-} &= \gamma\sin^2\theta\cos^2\theta, \\
    \Gamma_{+\rightarrow-} &= \gamma\cos^4\theta, \\
    \Gamma_{-\rightarrow+} &= \gamma\sin^4\theta,
\end{split}  \label{S16}
\end{align}
where $\gamma$ is the radiative decay rate of $\ket{e}\rightarrow \ket{1}$, and $\theta\equiv\arctan(-\Omega/\Delta)/2 + \pi H(\Delta)/2$ is the dressed-state mixing angle introduced in Eq.~\eqref{eq:dressed_state}. The transitions $\ket{\pm,N}\rightarrow \ket{\pm,N-1}$ correspond to elastic scattering at the driving frequency $\omega_L$ with a rate of $\Gamma_{\pm\rightarrow\pm}$. In contrast, $\Gamma_{\pm\rightarrow\mp}$ denotes inelastic scattering that contributes to the Mollow sidebands at frequencies $\omega_L\pm \sqrt{\Delta^2+\Omega^2}$.

We next express the dressed-state transition rates in terms of the saturation parameter $s\equiv\frac{2\Omega^2}{4\Delta^2+\gamma^2}=\frac{s_0}{4(\Delta/\gamma)^2+1}$ with $s_0=\frac{2\Omega^2}{\gamma^2}$, where negligible spatial dependence  has been omitted. In the far blue-detuned regime, $|\Delta|\gg\gamma$, and $\Delta>0$, 
\begin{equation} \label{S17}
    s \simeq \frac{\Omega^2}{2\Delta^2}=\frac{1}{2}\left(\frac{1}{\cos^22\theta}-1\right). \\
\end{equation}
Using Eq.~\eqref{S17}, the transition rates in Eq.~\eqref{S16} become
\begin{align}
\begin{split}
    \Gamma_1 &= \Gamma_{+\rightarrow+} = \Gamma_{-\rightarrow-} = \frac{\gamma}{2}\frac{s}{1+2s}, \\
    \Gamma_2 &= \Gamma_{+\rightarrow-} = \frac{\gamma}{2}\left(\frac{1+s}{1+2s} - \frac{1}{\sqrt{1+2s}} \right), \\
    \Gamma_3 &= \Gamma_{-\rightarrow+} = \frac{\gamma}{2}\left(\frac{1+s}{1+2s} + \frac{1}{\sqrt{1+2s}} \right).
\end{split}
\end{align}
The steady-state dressed-state populations are $\sigma_+^\text{st}=\Gamma_3/(\Gamma_2+\Gamma_3)$ and $\sigma_-^\text{st}=\Gamma_2/(\Gamma_2+\Gamma_3)$. In the weak-saturation limit $s\ll1$, $\Gamma_3 \gg \Gamma_2$, and the atom therefore occupies predominantly $\ket{+}$. For red detuning, $\Gamma_2$ and $\Gamma_3$ are interchanged.

Using the steady-state coherent-to-incoherent scattering ratio $1:s$~\cite{Tannoudji1998}, the collected fluorescence rates are
\begin{align}
\begin{split} \label{S19}
    R_\text{rc}(s) &= \frac{\eta_\text{c}\gamma}{2}\frac{s}{(1+s)(1+2s)}, \\
    R_\text{nrc}(s) &= \frac{\eta_\text{c}\gamma}{2}\frac{s^2}{(1+s)^2(1+2s)}, \\
    R_\text{ic}(s) &= \frac{\eta_\text{c}\gamma}{2}\frac{s^2}{(1+s)^2}.
\end{split}
\end{align}
Here, $R_\text{rc}(s)$, $R_\text{nrc}(s)$, and $R_\text{ic}(s)$ denote the resonant coherent, nonresonant coherent, and incoherent photon-collection rates, respectively. Their sum recovers the total photon-collection rate, $R_\text{rc}(s)+R_\text{nrc}(s)+R_\text{ic}(s)=\frac{\eta_\text{c}\gamma}{2}\frac{s}{1+s}=R(s)$.

\subsubsection{Heating mechanisms and verification of the motion-induced decoherence effect}\label{supp:heating_mechanisms}

The two heating mechanisms considered are photon recoil and dipole-force fluctuations (DFF). Photon recoil transfers momentum to the atom during photon-scattering events. Averaging over the stochastic atomic motion and the dipole-radiation pattern, the mean energy increase per scattering event is $2E_\text{R}$, where $E_\text{R}=(\hbar k_L)^2/(2m)$ is the recoil energy.

DFF arises from stochastic switching between the dressed-state potentials $\ket{+}$ and $\ket{-}$~\cite{Martinez2018,Tannoudji1985}. For blue-detuned light, the atom predominantly occupies $\ket{+}$, and transitions to $\ket{-}$ occur at a rate $\sigma_+^{\mathrm{st}}\Gamma_{+\rightarrow-}$. To evaluate the associated heating, we approximate the atomic motion as harmonic oscillation in each spatial direction. We define the temperature as $T\equiv \mathbb{E}[E]/(3k_\mathrm{B})$, where $\mathbb{E}[\cdot]$ denotes the ensemble average over atomic trajectories and $E$ is the total motional energy of one trajectory. The oscillation amplitude $\zeta_0$ along each direction ($\zeta=x,y,z$) follows the distribution derived from the Maxwell-Boltzmann distribution:
\begin{equation}
    f(\zeta_0) = \frac{\zeta_0}{\sigma_\zeta^2} \exp\left( -\frac{\zeta_0^2}{2\sigma^2_\zeta} \right), \\
\end{equation}
where $\sigma_\zeta=\sqrt{k_\mathrm{B}T/(m\omega_\zeta^2)}$. For a given amplitude $\zeta_0$, the probability density of the position $\zeta$ is
\begin{equation}
    g(\zeta) = \frac{1}{\pi\sqrt{\zeta_0^2-\zeta^2}}, \\
\end{equation}
and the corresponding momentum magnitude is
\begin{equation}
    p(\zeta) = m\omega_\zeta\sqrt{\zeta_0^2-\zeta^2}, \\
\end{equation}
where the momentum direction can be along either the $+\zeta$ or $-\zeta$ direction. A DFF event transfers the atom from the $\ket{+}$ potential to the $\ket{-}$ potential for a residence time $\tau_-$ before returning to $\ket{+}$. Assuming that the residence time $\tau_-$ follows an exponential distribution with decay rate $\Gamma_3$, the second moment is
\begin{equation}
    \int_0^\infty \tau_-^2 \Gamma_3 e^{-\Gamma_3\tau_-}\,d\tau_-
    =\frac{2}{\Gamma_3^2}.
\end{equation}

Averaging over the two momentum directions, the resulting increase in motional energy along the $\zeta$ axis is thus
\begin{align}
\begin{split}
    \Delta E_{\text{DFF},\zeta} &= \frac{1}{2}m\omega_\zeta^2 \left( \frac{1}{2}\left( \zeta - \frac{p(\zeta)\tau_-}{m} \right)^2 + \frac{1}{2}\left( \zeta + \frac{p(\zeta)\tau_-}{m} \right)^2 - \zeta^2 \right) \\
    &= \frac{\left(p(\zeta)\omega_\zeta\tau_-\right)^2}{2m}, \\
\end{split}
\end{align}
of which the average becomes
\begin{equation}
    \int_0^\infty \Delta E_{\text{DFF},\zeta}\Gamma_3e^{-\Gamma_3\tau_-}d\tau_- = \frac{p^2(\zeta)\omega_{\zeta}^2}{m\Gamma_3^2},
\end{equation}
and the total average energy increase per DFF event $\Delta E_\text{DFF}$ is
\begin{equation}
    \Delta E_\text{DFF} = \sum_{\zeta=x,y,z} \left[ \int_0^{\infty} d\zeta_0\int_{-\zeta_0}^{\zeta_0}d\zeta ~ \frac{p^2(\zeta)\omega_\zeta^2}{m\Gamma_3^2}f(\zeta_0)g(\zeta) \right]. \\
\end{equation}
Including both photon recoil and DFF, the ensemble-averaged motional energy follows
\begin{align}
\begin{split}
    \frac{d\mathbb{E}[E]}{dt} &= E_\text{R}\frac{\gamma s}{1+s} + \mathbb{E}[E]\frac{\Gamma_2\tilde{\omega}^2}{\Gamma_3(\Gamma_2+\Gamma_3)}, \\
\end{split} \label{eq:S25}
\end{align}
where $\tilde{\omega}^2=(\omega_x^2+\omega_y^2+\omega_z^2)/3$. The first and second terms describe heating from photon recoil and DFF, respectively. 

\begin{figure}
\centering
\includegraphics[width=0.90\textwidth]{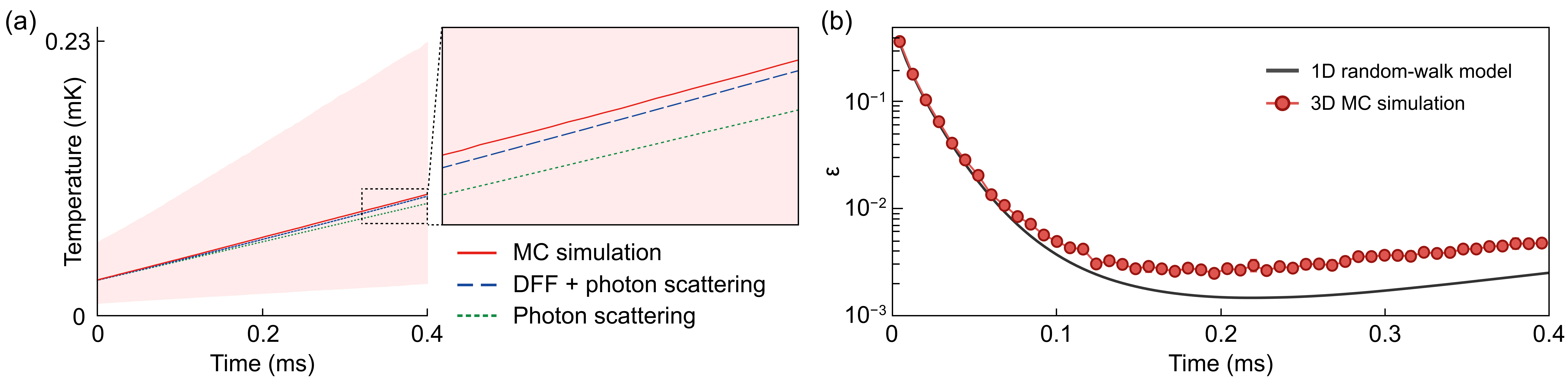} 
\caption{
\textbf{1D random-walk model.}
    \textbf{(a)}
    Atomic temperature ($T \equiv \mathbb{E} [E]/3k_\mathrm{B}$) as a function of detection time for an atom in a 7 mK trap depth with $s=0.08$. The red, blue, and green curves represent the Monte Carlo simulation, the analytical model including dipole-force fluctuations (DFF), and the analytical model including photon-recoil heating only, respectively. The shaded regions indicate the 90\% confidence intervals.
    \textbf{(b)}
    Comparison of the 3D Monte Carlo simulation and the 1D amplitude random-walk model for the detection error. Error bars indicate one standard deviation across 10 independent Monte Carlo ensembles, each containing 5,000 trajectories.
}
\label{fig:S1}
\end{figure}

The temperature predicted by Eq.~\eqref{eq:S25} agrees with the Monte Carlo simulations (Fig.~\ref{fig:S1}(a)). Under the conditions considered, DFF heating is negligible compared with photon-recoil heating and makes only a small contribution to motion-induced decoherence. DFF can become significant when $\tilde{\omega}/\gamma$ exceeds the Lamb-Dicke parameter~\cite{Karan2024} or when the driving field approaches resonance~\cite{Martinez2018}, neither of which applies here.

To assess the dominant motional contribution to the BHD decoherence, we compare a one-dimensional random-walk model of the axial motion with the full three-dimensional Monte Carlo simulations. Photon recoils are modeled as a random walk in momentum space, with the axial trajectory described by $z=z_0(t)\sin(\omega_zt-\phi_z(t))$, where $z_0(t)$ and  $\phi_z(t)$ evolve under recoil-induced heating. For each scattering event, the absorption recoil is applied along the propagation direction of SDB, while the emission direction is sampled from the $\sigma^+$ polarization dipole-radiation pattern. Figure~\ref{fig:S1}(b) shows that the 1D random-walk model reproduces the time dependence of the detection error, including an optimal detection time due to the competition between coherent-signal accumulation and recoil-induced motional decoherence. The residual discrepancy reflects additional motional effects in the full 3D simulations, as discussed in the main text, which reduce the mean BHD signal and increase its variance.

\subsubsection{Characteristics of the state-detection beam}\label{supp:SDB}
\begin{figure}
\centering
\includegraphics[width=0.90\textwidth]{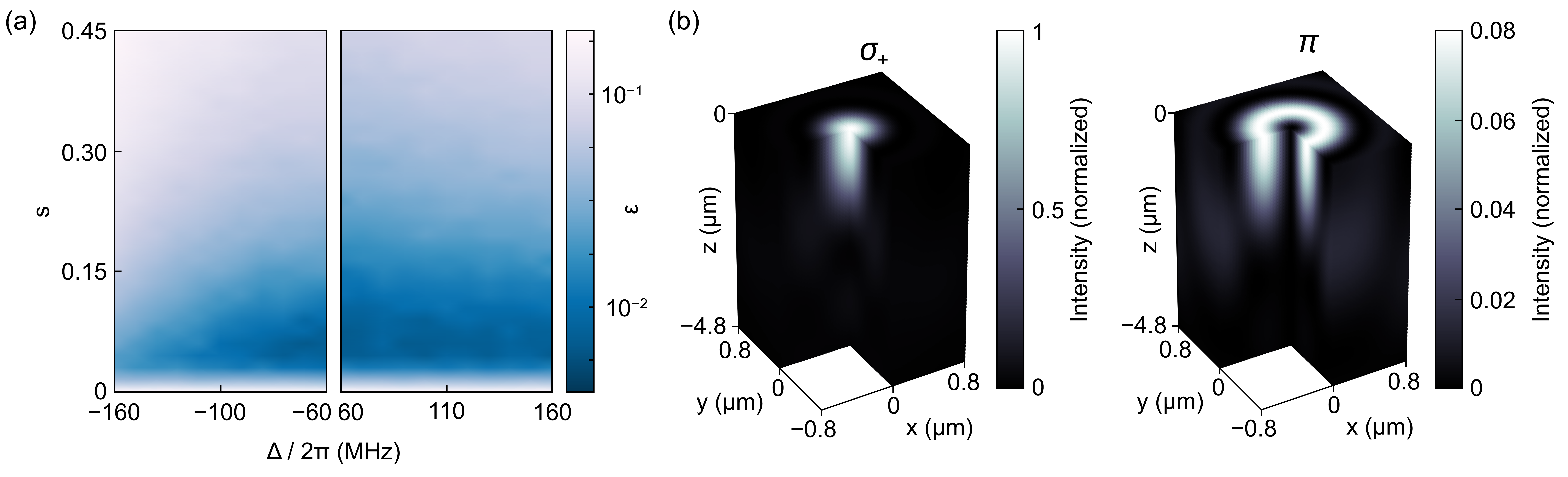} 
\caption{ 
    \textbf{Characteristics of the state-detection beam.}
    \textbf{(a)} Detection error as a function of the saturation parameter $s$ and detuning $\Delta$ of the SDB for a trap depth of 7 mK and a detection time of 200~$\upmu\mathrm{s}$.
    \textbf{(b)} Spatial intensity distribution of $\sigma_+$ and $\pi$ polarization components for a pure $\sigma_+$ input with an SDB $1/e^2$ intensity radius of 0.35~$\upmu\mathrm{m}$. Both intensities are normalized to the peak $\sigma_+$ intensity. The top surface $(z=0)$ corresponds to the atomic plane perpendicular to the SDB propagation direction.
    }
\label{fig:S2}
\end{figure}

We first consider an ideal plane wave driving a closed cycling transition between the bright state and the excited state. The quantization axis is defined by a $9.8~\mathrm{G}$ bias magnetic field aligned with the SDB propagation axis. The SDB is purely $\sigma_+$-polarized with respect to this quantization axis. We evaluate the BHD detection error as a function of the saturation parameter $s$ and detuning $\Delta$ of the SDB. As shown in Fig.~\ref{fig:S2}(a), the detection fidelity depends only weakly on $|\Delta|$ in the far-detuned regime, but is slightly lower for red detuning than for blue detuning of the same magnitude. This asymmetry results from position-dependent dressed-state energy shifts that bring the transition closer to resonance at certain positions within the optical tweezer.

We next extend the model to a tightly focused Gaussian SDB for site-selective readout. The full vectorial field is calculated, including the spatial distribution of its polarization components (Fig.~\ref{fig:S2}(b)). In the strongly focused regime, breakdown of the paraxial approximation generates longitudinal field components and polarization mixing. These unwanted polarization components can drive transitions to neighboring Zeeman sublevels, producing population leakage out of the cycling manifold and reducing the detection fidelity.

\subsubsection{Simulation procedure}\label{supp:simulation_procedure}

Unless otherwise specified, the Monte Carlo simulations use the parameters summarized in Table~\ref{table:simulation_parameters}.

\begin{table}[htbp]
\renewcommand{\arraystretch}{1.25}
\centering
\begin{tabular}{
    m{0.38\textwidth}
    m{0.18\textwidth}
    m{0.24\textwidth}}
    \hline
    \hline
    Parameter & Symbol & Value \\
    \hline
    SDB wavelength & $\lambda_L$ & $780~\mathrm{nm}$ \\
    Radiative decay rate & $\gamma$ & $2\pi\times6.07~\mathrm{MHz}$ \\
    Tweezer wavelength & $\lambda_{\mathrm{tw}}$ & $860~\mathrm{nm}$ \\
    Tweezer waist & $W$ & $0.516~\upmu\mathrm{m}$ \\
    Tweezer Rayleigh range & $z_{\mathrm R}$ & $0.526~\upmu\mathrm{m}$ \\
    Trap depth & $V_0/k_{\mathrm B}$ & $7~\mathrm{mK}$ \\
    Radial trap frequency & $\omega_r/2\pi$ & $509.1~\mathrm{kHz}$ \\
    Axial trap frequency & $\omega_0/2\pi$ & $352.9~\mathrm{kHz}$ \\
    Initial temperature & $T_0$ & $30~\upmu\mathrm{K}$ \\
    Center detuning & $\Delta/2\pi$ & $123~\mathrm{MHz}$ \\
    Center saturation parameter & $s$ & $0.08$ \\
    \hline
    \hline
\end{tabular}
\caption{Parameters used in the Monte Carlo simulations.}
\label{table:simulation_parameters}
\end{table}

For a far-detuned detection beam, the Monte Carlo simulation proceeds as follows~\cite{Martinez2018,Zipkes2011}:

\begin{enumerate}
    \item Initialize each trajectory in the $\ket{+}$ dressed state and sample the atomic position and velocity from the Maxwell--Boltzmann distribution at temperature $T_0$.
    
    \item Calculate the allowed transition rates, including $\Gamma_{\pm\rightarrow\pm}$, $\Gamma_{\pm\rightarrow\mp}$, and scattering processes among 
    \begin{equation*}
        \left\{\ket{5S_{1/2},F=2,m_F=2}, \ket{5S_{1/2},F=2,m_F=1}, \ket{5S_{1/2},F=1,m_F=1}\right\}
    \end{equation*}
    and
    \begin{equation*}
        \left\{\ket{5P_{3/2},F'=3,m_F'=3}, \ket{5P_{3/2},F'=3,m_F'=2}, \ket{5P_{3/2},F'=2,m_F'=2}\right\},
    \end{equation*}
    including polarization-mixing effects. For each transition $i$,  determine the maximum rate $\Gamma_{i,\text{max}}$ over all accessible positions.
    
    \item For each transition, sample a candidate waiting time from  $P(t)=\Gamma_{i,\text{max}}e^{-\Gamma_{i,\text{max}}t}$ and select the shortest time $\tau$.
    
    \item Propagate the atom classically for time $\tau$ in the potential corresponding to its current dressed state, $\ket{+}$ or $\ket{-}$. The optical tweezer potential for $\ket{1}$ is
    \begin{equation}
        V(x,y,z) = \frac{-V_0}{1+(z/z_\text{R})^2}\exp\left(\frac{-2(x^2+y^2)/W^2}{1+(z/z_\text{R})^2}\right),
    \end{equation}
    where $V_0$ is the trap depth, $W$ is the beam waist, and $z_\text{R}$ is the Rayleigh range. At the updated position, evaluate the transition rate $\Gamma_i$ and accept the candidate transition with probability $\Gamma_i/{\Gamma_{i,\text{max}}}$. 
    
    \item If the transition is accepted, apply the momentum change associated with directed photon absorption and spontaneous emission, with the emission direction sampled from the corresponding dipole-radiation pattern. If the transition is rejected, retain the evolved position and momentum and return to step (2).

    \item During the trajectory evolution, calculate the coherent photon-collection rate $R_\text{rc}$ from Eq.~\eqref{S19} and the motion-induced phase shift $\phi_\text{osc}(t)=2k_Lz(t)$. These quantities are used in Eq.~\eqref{eq:BHD_signal} of the main text to accumulate the BHD signal over the interval $\tau$.

    \item Terminate the trajectory when (i) the total evolution time reaches the detection time $t_\text{d}$, (ii) the total mechanical energy becomes nonnegative, in which case the atom is classified as lost from the optical tweezer, or (iii) the atom decays into the optically dark $\ket{0}$.

\end{enumerate}

\subsection{Classical optical communication} \label{supp:classical}
We compare photon-number-resolving (PNR) and coherent detection schemes in terms of their signal-to-noise ratio (SNR) and bit error rate (BER). We first consider lossless PNR detection with no background counts. The bright-state ($\ket{1}$) photon number follows a Poisson distribution with mean photon number $N$, while the optically dark $\ket{0}$ produces zero photons. Since an error occurs only when no photon is detected from the bright state, the corresponding BER becomes $e^{-N}/2$. 

The SNR and BER of amplitude-shift keying (ASK) provide a useful reference for the coherent fluorescence readout considered here. The dark and bright states correspond to zero and nonzero coherent field amplitudes, respectively, analogous to binary ASK. For ideal homodyne detection, the resulting BER is $Q(\sqrt{N_\text{rc}})$, where $N_\text{rc}$ is the detected coherent-photon number and $Q(x)=1-\Phi(x)$, with $\Phi(x)$ denoting the cumulative distribution function of the standard normal distribution. For comparison, the two-channel binary phase-shift keying (BPSK) scheme assigns opposite fluorescence phases to the two qubit states. This doubles the quadrature-signal separation, while independent shot noise from the two local oscillators doubles the noise variance, resulting in a BER of $Q(\sqrt{2N_\text{rc}})$. The BPSK scheme therefore requires half the detected coherent-photon number of the ASK-like scheme to achieve the same BER. 

Table~\ref{table1} summarizes the noiseless performance of PNR, ASK and BPSK detection. We next include detector noise to compare coherent fluorescence readout with realistic photon counting and camera-based detection. For BHD, the main additional limitation considered in the main text is motion-induced decoherence.

\begin{table} [htbp]
\renewcommand{\arraystretch}{1.35}
\centering
\begin{tabular}{m{0.18\textwidth}<{\centering}||m{0.18\textwidth}<{\centering}|m{0.18\textwidth}<{\centering}|m{0.18\textwidth}<{\centering}}
     \hline
     Method & SNR & BER & $N (N_\text{rc}{})$ for same BER \\ 
     \hline
     PNR  & $\sqrt{N}$ (0 dB)    &  $\sim\frac{1}{2}e^{-SNR^2}$ ($0~\text{dB}$)  & $N$ \\ 
     \hline
     Coherent (ASK)  &   $2\sqrt{N_\text{rc}}$ (+3 dB)  &  \multirow{2}{*}{$Q(SNR/2)$($\sim-4 ~\text{dB}$)} & $\sim 1.58\times N$ \\ 
     \cline{1-2}\cline{4-4}
     Coherent (BPSK) &    $2\sqrt{2N_\text{rc}}$ (+6 dB) &  & $\sim0.79\times N$ \\ 
     \hline
\end{tabular}
\caption{Comparison of ideal detection methods with noiseless detectors.
}
\label{table1}
\renewcommand{\arraystretch}{1}
\end{table}

To model photon-counting readout, we denote the dark- and bright-state signal distributions by $p_\text{d}(t)$ and $p_\text{b}(t)$, respectively, for a detection time $t$. For an ideal PNR detector with incident photon rate $\Gamma_\text{in}$ and background count rate $\Gamma_\text{b}$, these distributions are $p_\text{d}(t) = \mathcal{P}(\Gamma_\text{b}t)$ and $p_\text{b}(t) = \mathcal{P}((\Gamma_\text{in}+\Gamma_\text{b})t)$, where $\mathcal{P}(\lambda)$ denotes the Poisson distribution with mean $\lambda$. For CCD and CMOS cameras, we additionally include finite quantum efficiency, dark current, and readout noise. With quantum efficiency $\eta_\text{QE}$, dark-current rate $\Gamma_\text{d}$, and readout noise of $\sigma_\text{r}$, the dark- and bright-state distributions become
\begin{equation}
\begin{array}{r@{}l}
    p_\text{d}(t) &{}= \mathcal{P}((\eta_\text{QE}\Gamma_\text{b}+\Gamma_\text{d})t) \ast \mathcal{N}(\sigma_\text{r}), \\
    \\
    p_\text{b}(t) &{}= \mathcal{P}((\eta_\text{QE}(\Gamma_\text{in}+\Gamma_\text{b})+\Gamma_\text{d})t) \ast \mathcal{N}(\sigma_\text{r}),
\end{array}
\end{equation}
where $\ast$ denotes convolution and $\mathcal{N}(\sigma)$ is a zero-mean normal distribution with standard deviation $\sigma$.

We further model electron-multiplying CCDs (EMCCDs) and avalanche photodiodes (APDs) using the stochastic framework of Ref.~\cite{Hirsch2013}. For EMCCDs, clock-induced charge (CIC), characterized by the mean charge $\lambda_\text{CIC}$, contributes additional background electrons. Electron multiplication with gain $g$ is described by
\begin{equation} \label{eqS2}
    {\mathcal G}(n_\text{out};n_\text{in},g) = \frac{n_\text{out}^{n_\text{in}-1}e^{-n_\text{out}/g}}{\Gamma(n_\text{in})g^{n_\text{in}}}, 
\end{equation}
which gives the probability density for $n_\text{out}$ output electrons conditional on $n_\text{in}$ input electrons. Here, $\Gamma(x)$ denotes the gamma function, and ${\mathcal G}$ is equivalent to the probability density function of a gamma distribution $f(x;k,\theta)=\frac{x^{k-1}e^{-x/\theta}}{\Gamma(k)\theta^k}$ with shape parameter $k=n_\text{in}$ and scale parameter $\theta=g$. Including clock-induced charge, multiplication and readout noise, the final signal distributions are
\begin{equation}
\begin{array}{r@{}l}
\hspace*{-0.1cm} p_\text{d}(t) &{}= \left\{\mathcal{P}(n_\text{in}=\eta_\text{QE}\Gamma_\text{b}t+\Gamma_\text{d}t+\lambda_\text{CIC}) \ast {\mathcal G}(n_\text{out};n_\text{in},g)\right\} \ast \mathcal{N}(\sigma_\text{r}), \\
 \\
\hspace*{-0.1cm} p_\text{b}(t) &{}= \left\{\mathcal{P}(n_\text{in}=\eta_\text{QE}(\Gamma_\text{in}+\Gamma_\text{b})t+\Gamma_\text{d}t+\lambda_\text{CIC}) \ast {\mathcal G}(n_\text{out};n_\text{in},g)\right\} \ast \mathcal{N}(\sigma_\text{r}).
\end{array}
\end{equation}
The detector parameters used for modeled sCMOS and APD/EMCCD performance in Fig.~\ref{fig:6} are summarized in Table~\ref{table2}.

\begin{table} [htbp]
\renewcommand{\arraystretch}{1.2}
\centering
\begin{tabular}{m{0.14\textwidth}<{\centering}m{0.1\textwidth}<{\centering}||m{0.16\textwidth}<{\centering}|m{0.16\textwidth}<{\centering}|m{0.16\textwidth}<{\centering}|m{0.08\textwidth}<{\centering}|m{0.08\textwidth}<{\centering}}
     \hline
     \multicolumn{2}{c||}{\multirow{2}{*}{}} & Readout noise   & Background noise   & Quantum efficiency   & CIC & Gain  \\
     & & $\sigma_\text{r}~(e^-)$ & $\Gamma_\text{b}~(\text{photon}/\mathrm{ms})$ & $\eta_\text{QE}$  & $\lambda_\text{CIC}~(e^-)$ & $g$ \\
     \hline
     \multicolumn{2}{c||}{sCMOS} &  0.27  & 0.15 & 0.9 & - & - \\ 
     \hline
     \multirow{2}{*}{APD/EMCCD} & Noisy &  10  & 1.5 & 0.5 & 1 & 30 \\ 
     \cline{2-7}
      & Low $\Gamma_\text{b}$ &  10  & 0.015  & 0.95 & 0 & 30 \\ 
     \hline
\end{tabular}
\caption{Noise parameters used in Fig.~\ref{fig:6}.}

\label{table2}
\renewcommand{\arraystretch}{1}
\end{table}

\end{document}